\documentclass[%
floatfix,
reprint,
amsmath,amssymb,
aps, pre,
superscriptaddress
]{revtex4-2}
\usepackage[english]{babel}

\usepackage{amsfonts,array}
\usepackage{mathtools}
\usepackage{bm}
\usepackage{dcolumn}
\usepackage[output-decimal-marker={.}]{siunitx} 

\usepackage[final]{graphicx} 
\usepackage{listings} 
\usepackage[usenames]{color} 

\usepackage{comment} 
\usepackage{tikz} 

\usepackage{color}
\usepackage{float}

\definecolor{xblue}{rgb}{0.2,0.2,1.0}

\usepackage{hyperref}
\hypersetup{
	colorlinks=true,
	linkcolor={blue}, 
	filecolor=magenta,   
	citecolor={blue}, 
	urlcolor=blue, 
}

\date{\today}

\begin{document}
	
	
\title{Geometry-controlled active transport in smectic textures}
	
    \author{Yhony M. Arce}
    \email{ym.arce@unesp.br}
    \affiliation{Instituto de F\'isica Te\'orica, Universidade Estadual Paulista, S\~ao Paulo, SP, Brazil}
    \author{William G. C. Oropesa}
    \affiliation{Instituto de F\'isica Te\'orica, Universidade Estadual Paulista, S\~ao Paulo, SP, Brazil}
    \affiliation{ICTP South American Institute for Fundamental Research, S\~ao Paulo, SP, Brazil}
    \author{Alef R. Siqueira}
    \affiliation{Instituto de F\'isica Te\'orica, Universidade Estadual Paulista, S\~ao Paulo, SP, Brazil}
    \author{Hartmut Löwen}
    \affiliation{Institut für Theoretische Physik II: Weiche Materie, Heinrich-Heine-Universität Düsseldorf, Düsseldorf, Germany}
    \author{André P. Vieira}
    \affiliation{Instituto de F\'isica, Universidade de São Paulo, S\~ao Paulo, SP, Brazil}
    \author{Danilo B. Liarte}
    \email{danilo.liarte@unesp.br}
    \affiliation{Instituto de F\'isica Te\'orica, Universidade Estadual Paulista, S\~ao Paulo, SP, Brazil}
    \affiliation{ICTP South American Institute for Fundamental Research, S\~ao Paulo, SP, Brazil}
	
\begin{abstract}
Self-propelled particles moving through structured anisotropic media such as smectic liquid crystals exhibit transport properties that differ qualitatively from those in isotropic environments.
Here we introduce a minimal model of an active Brownian particle whose propulsion axis is coupled to the layer-normal field of a smectic liquid crystal.
Using numerical simulations complemented by approximate analytical calculations, we investigate the dynamics of an active particle in three representative two-dimensional smectic microstructures: a uniform planar texture, an isolated toroidal focal conic domain, and a periodic array of toroidal focal conic domains.
For the planar and isolated focal-conic geometries, we characterize the crossover from short-time superdiffusion to long-time normal diffusion.
In periodic focal-conic arrays, however, transport is controlled by two distinct persistence mechanisms whose relative importance depends on the Péclet number ($\mathrm{Pe}$) and the coupling strength $\varepsilon$ between the particle orientation and the smectic layer-normal field.
At small $\varepsilon$, long-time diffusion is governed by the intrinsic persistence time of the active Brownian dynamics.
At large $\varepsilon$, repeated scattering by smectic defects generates an emergent persistence time that dominates the transport.
These results demonstrate that the competition between the intrinsic persistence of active motion and the mean free time between encounters with smectic defects provides a simple physical mechanism for the emergence of nontrivial diffusive behavior in structured anisotropic media.
\end{abstract}

	\maketitle

\section{Introduction}

Active matter~\cite{ramaswamy2010mechanics,bechinger2016active,te-VrugtCates2026} consists of self-driven microscopic units that continuously consume energy to sustain autonomous motion, giving rise to transport phenomena with no counterpart in equilibrium systems. Such transport effects in nonequilibrium active matter are often governed by the competition between distinct dynamical time scales.
One of the simplest examples for a self-propelled agent is an active Brownian particle (ABP)~\cite{romanczuk2012active}, whose dynamics are characterized by an intrinsic persistence time $\tau_p = D_R^{-1}$ associated with the rotational diffusion coefficient $D_R$.
This persistence time determines the crossover from short-time superdiffusive (often ballistic) motion to long-time normal diffusion and sets the scale of the effective long-time diffusivity.
Additional physical processes, including interactions with other particles, confining boundaries, or heterogeneous environments, introduce new dynamical time scales that can profoundly modify transport, giving rise to phenomena such as clustering~\cite{cates2015motility, Maggi2021, oropesaLiarte2026}, wall accumulation~\cite{wall1,wall2,wall3}, anomalous diffusion~\cite{dieterich2008anomalous, Angelico2006}, and trapping and jamming in disordered landscapes \cite{Reichhardt2014,DebetsJanssen2022}.
Indeed, confinement and spatial structuring of the environment are increasingly recognized as a means of steering the motion and self-organization of active matter, rather than merely as a source of perturbations (see, e.g., Ref. \cite{Araujo2023} and references therein); periodic arrays of obstacles, for instance, have been shown to organize the mesoscale turbulence of dense bacterial suspensions into regular vortex lattices \cite{Reinken2020}.
In this work we investigate how complex structured environments generate an emergent persistence mechanism that competes with the intrinsic orientational persistence of active Brownian motion, leading to qualitatively new transport behavior in heterogeneous environments.

Liquid crystals~\cite{de1993physics,chaikin1995principles} provide a paradigmatic class of experimentally accessible structured environments in which the anisotropy of the host medium strongly influences particle transport~\cite{Lavrentovich2014}.
Even relatively simple liquid-crystalline textures give rise to transport behavior that differs qualitatively from that observed in isotropic fluids~\cite{Lavrentovich2016}.
For example, while active Brownian particles in isotropic environments undergo conventional diffusion at long times, active particles moving in nematic liquid crystals exhibit anomalous transport, with ballistic motion parallel to the director and logarithmically enhanced superdiffusion in the transverse direction~\cite{Toner2016}.
Similar anisotropic transport has also been observed for externally driven colloids dispersed in nematic liquid crystals, where the distortions induced by the inclusions further influence the dynamics~\cite{Pages2019}.
In smectic liquid crystals, even the simplest texture, consisting of parallel layers, exhibits anomalous transverse transport arising from long-wavelength fluctuations of the layer-normal field after a characteristic crossover time~\cite{Ferreiro2018}.
Here, the anomalous dynamics originates from thermal fluctuations of an otherwise homogeneous texture.
A natural question is how this picture is modified when transport is instead governed by repeated encounters with the complex geometry of focal-conic textures.

Focal conic domains are among the most characteristic microstructures of smectic liquid crystals.
Their remarkable geometry played a central role in the discovery of the smectic layered structure, as the observation of well-defined ellipses and hyperbolas in early microscopy experiments led Friedel and Grandjean to infer that the layers form families of equidistant Dupin cyclides~\cite{Friedel1910a,Friedel1910b,GraySal2006}.
These surfaces fill space while remaining singular along pairs of focal conics, which may consist of an ellipse and a hyperbola, a circle and a straight line, or two parabolas~\cite{kleman2003ch10}.
More recently, the constraint of constant layer spacing has been shown to endow smectic textures with a hidden Poincaré symmetry~\cite{AlexanderKamien2010}, providing a key ingredient for a unified geometric framework for understanding the organization of focal conic domains~\cite{LiarteSethna2016}.
Advances in both experiments~\cite{MeyerFoy2009,HonglawanYang2013,GharbiSte2015,LiarteSethna2015,Wittmann2021,Monderkamp2021} and numerical simulations~\cite{LiarteSethna2015,Wittmann2021,Monderkamp2021} have revealed a remarkable diversity of smectic textures in which focal conic domains emerge spontaneously and organize into complex microstructures.
These geometric features naturally raise the question of how focal-conic textures influence active transport.
If the propulsion direction of an active particle tends to align with the local layer-normal field, the curved smectic geometry steers the particle trajectories toward focal conic defects, while repeated encounters with these defects act as scattering events that reorient the particle.
As we show below, the competition between this geometry-induced persistence and the intrinsic orientational persistence of active Brownian motion gives rise to qualitatively new transport regimes.

In this work, we introduce a minimal theoretical model for the motion of active particles through structured smectic-A textures.
The dynamics are described by coupled translational and rotational Langevin equations that account for both self-propulsion and alignment with the local smectic layer-normal field.
The model is governed primarily by the Péclet number, $\mathrm{Pe}$, which quantifies the activity, and by an anchoring parameter, $\varepsilon$, that controls the coupling between the particle orientation and the smectic texture.
We investigate transport in three representative two-dimensional cross sections of increasing geometric complexity: a uniform planar texture, an isolated toroidal focal conic domain, and a periodic array of toroidal focal conic domains.
For the planar geometry, we show that coupling to the layer-normal field renormalizes the intrinsic persistence of active Brownian motion, extending the ballistic regime and enhancing the effective diffusivity.
For an isolated focal conic domain, we identify the elementary scattering processes associated with the line and point singularities of the smectic texture.
Our central result concerns periodic focal-conic arrays, where repeated scattering generates a geometry-induced persistence mechanism that qualitatively changes the long-time transport.
When the intrinsic persistence dominates, the effective diffusivity follows the conventional active Brownian scaling, $D_{\mathrm{eff}}\sim\mathrm{Pe}^2$.
In contrast, when the geometry-induced persistence becomes dominant, transport is controlled by the mean free time between successive encounters with focal conic defects, leading to the distinct scaling $D_{\mathrm{eff}}\sim\mathrm{Pe}$.
These results demonstrate how the competition between intrinsic and environment-induced persistence governs active transport through structured smectic textures.

The remainder of this paper is organized as follows.
In Sec.~\ref{sec:Model}, we introduce the theoretical model and the numerical methods.
Sections~\ref{sec:Planar}, \ref{sec:OneFCD}, and \ref{sec:ArrayFCDs} present the analytical and numerical results for the planar texture, an isolated focal conic domain, and a periodic array of focal conic domains, respectively. 
Finally, Sec.~\ref{sec:Conclusions} summarizes our main findings and discusses their broader implications.

\section{Model and simulations}
\label{sec:Model}

\begin{figure}[!h]
\begin{center}
\includegraphics[width=1.0\linewidth]{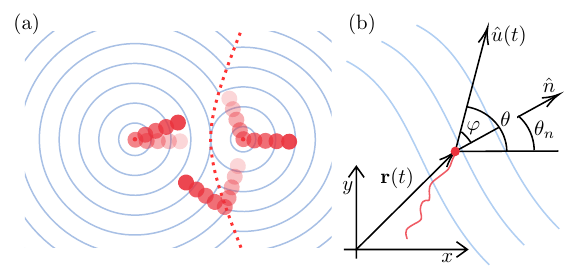}
\end{center}
\vspace{-0.6cm}
\caption{(a) Illustration of an active particle moving in the layered configurations of a Sm-A phase.
Particles (red disks) tend to move in a direction perpendicular to the layers (blue lines) and can be scattered by defects (red dashed lines).
(b) Illustration of the main vector and angle definitions.}
\label{fig:Illustration}
\end{figure}

We consider the overdamped dynamics of a single active Brownian particle moving through a prescribed two-dimensional smectic-A texture (Fig.~\ref{fig:Illustration}).
The instantaneous state of the particle is specified by its position $\bm{r}(t)$ and the unit vector $\hat{\bm{u}}(t)=(\cos\theta,\sin\theta)$ describing the direction of self-propulsion [Fig.~\ref{fig:Illustration}(b)].
Neglecting translational Brownian diffusion, the dynamics are governed by the coupled Langevin equations
\begin{equation}
\frac{d\bm{r}(t)}{dt}=v\,\hat{\bm{u}}(t),
\label{eq:tem}
\end{equation}
\begin{equation}
\frac{d\theta}{dt}
=
-\frac{\delta E}{\delta\theta}
+\eta(t),
\label{eq:theta-evo-2D}
\end{equation}
where $v$ is the self-propulsion speed and $\eta(t)$ is a Gaussian white noise with zero mean and variance $\langle\eta(t)\eta(t')\rangle = 2D_R\delta(t-t')$, which defines the rotational diffusion coefficient $D_R$.
The deterministic contribution to the angular dynamics arises from an effective alignment potential $E$ that couples the propulsion direction $\hat{\bm{u}}$ to the local smectic layer-normal field $\hat{\bm{n}}(\bm{r})$.
Respecting the nematic symmetry of the smectic phase, $\hat{\bm{n}}\leftrightarrow-\hat{\bm{n}}$, we adopt the simple form
\begin{equation}
E
=
-\frac{k}{2}
(\hat{\bm{u}}\cdot\hat{\bm{n}})^2,
\label{eq:EnergyAlignment}
\end{equation}
where $k$ controls the strength of the alignment between the particle orientation and the smectic layer-normal field.

Several simplifying assumptions underlie the model introduced above. First, Eq.~\eqref{eq:tem} does not include translational Brownian diffusion.
This simplification allows us to isolate the effects of the interplay between intrinsic orientational persistence and the geometry of the smectic texture on active transport.
Since the nontrivial dynamical regimes identified below originate from the competition between characteristic orientational time scales, translational diffusion is not expected to qualitatively modify the physical mechanisms reported here over the time scales of interest.
Second, we treat the smectic layer-normal field $\hat{\bm{n}}(\bm{r})$ as a prescribed, time-independent background.
This assumption distinguishes the present work from that of Ferreiro-Córdova \emph{et al.}~\cite{Ferreiro2018}, where anomalous transport arises from long-wavelength space-time fluctuations of an otherwise homogeneous smectic texture.
Here, by contrast, the nontrivial dynamics originates from the geometry of the smectic microstructure itself, which introduces a competing characteristic time scale through repeated encounters with focal-conic domains. 
Treating the smectic texture as fixed also neglects the back-action of the active particle on the host medium, including hydrodynamic interactions and distortions of the smectic layers induced by surface anchoring.
This approximation is appropriate when the elastic energy required to deform the smectic texture is large compared with the active stresses generated by the swimmer.
Finally, the alignment potential in Eq.~\eqref{eq:EnergyAlignment} favors the propulsion direction to align parallel to the local layer-normal field.
This choice is intended to represent anisotropic active particles whose propulsion axis is coupled to their principal geometric axis and which preferentially align with the local smectic director through their shape and surface interactions with the surrounding mesogens.
The precise microscopic realization of this effective alignment depends on the particle geometry, surface anchoring, and the relative size of the active particle and the smectic layer spacing.
A complementary scenario, in which self-propulsion preferentially occurs along the layer tangent, would lead to qualitatively different transport properties and will be investigated elsewhere.

It is convenient to rewrite the equations of motion in dimensionless form.
Let $\theta_n(\bm{r})$ denote the angle between the local layer-normal field and a fixed reference axis [Fig.~\ref{fig:Illustration}(b)], and define the relative orientation $\varphi(t)=\theta(t)-\theta_n(\bm{r})$.
The alignment potential then becomes $E=-(k/2)\cos^2\varphi$, so that the angular dynamics is given by
\begin{equation}
    \frac{d\theta}{dt} = -\frac{k}{2}\sin(2\varphi) +\eta(t).
\label{eq:theta-evo-2D}
\end{equation}
Next, we measure time in units of the intrinsic rotational persistence time, $\tau_R=D_R^{-1}$, and length in units of a characteristic length scale $L$ associated with the smectic texture. Under the rescaling $t\rightarrow\tau_R t$ and $\bm{r}\rightarrow L\,\bm{r}$, the equations of motion become
\begin{equation}
    \frac{d\bm{r}}{dt} = \mathrm{Pe} \, \hat{\bm{u}},
\label{eq:em-tpara}
\end{equation}
\begin{equation}
    \frac{d\theta}{dt} = -\frac{\varepsilon}{2}\sin(2\varphi) + \sqrt{2}\,\xi(t),
\label{eq:p-aem}
\end{equation}
where $\xi(t)$ is a normalized Gaussian white noise with zero mean and variance $\langle\xi(t)\xi(t')\rangle=\delta(t-t')$.
The dynamics is therefore controlled by two dimensionless parameters. The first is the alignment strength
\begin{equation}
    \varepsilon = \frac{k}{D_R},
\label{eq:epsilonDefinition}
\end{equation}
which measures the ratio between the alignment torque and the rotational diffusion.
The second is the Péclet number,
\begin{equation}
    \mathrm{Pe} = \frac{l_p}{L} = \frac{v\tau_R}{L},
\label{eq:PecletDefinition}
\end{equation}
where $l_p=v\tau_R$ is the intrinsic persistence length of the active particle.
Throughout the remainder of this work we use the dimensionless variables but retain the original notation for simplicity.
For planar textures we choose for simplicity $L=l_p$, so that the dynamics depends only on the alignment parameter $\varepsilon$.
For focal-conic textures, $L$ is taken to be the characteristic size of the focal conic domain, and the transport properties are controlled by both $\mathrm{Pe}$ and $\varepsilon$.

The dimensionless Langevin equations, Eqs.~\eqref{eq:em-tpara} and~\eqref{eq:p-aem}, are integrated numerically using the Euler--Maruyama scheme, which is appropriate for stochastic differential equations with additive Gaussian noise.
Unless otherwise stated, we use a time step $\Delta t=0.001$ and integrate each trajectory up to a final time $T=10^6$.
Ensemble averages are computed over $N=5000$ independent trajectories for each set of model parameters.
For a generic observable $A(t)$, the ensemble average is defined as
\begin{equation}
    \langle A(t)\rangle = \frac{1}{N} \sum_{k=1}^{N} A_k(t).
\end{equation}
The chosen integration time and ensemble size ensure accurate estimates of the dynamical observables reported below, including the mean-square displacement and the rotational mean-square displacement.
The specific form of the smectic layer-normal field $\hat{\bm{n}}(\bm{r})$ depends on the texture under consideration and will be introduced separately for each of the three geometries studied below.

Throughout this work, transport is characterized by the mean-squared displacement (MSD),
\begin{equation}
    \mathrm{MSD}(t) = \left\langle |\bm{r}(t)-\bm{r}(0)|^2 \right\rangle,
    \label{eq:msdDef}
\end{equation}
and the rotational mean-squared displacement (RMSD),
\begin{equation}
    \mathrm{RMSD}(t) = \left\langle [\Delta\theta(t)]^2 \right\rangle.
    \label{eq:rmsdDef}
\end{equation}
Assuming that the orientational dynamics has reached a stationary state, the translational equation of motion, Eq.~\eqref{eq:em-tpara}, can be integrated to give
\begin{align}
    \mathrm{MSD}(t) &= \mathrm{Pe}^{2} \int_0^t \int_0^t C(|t'-t''|) \,dt'\,dt''
    \nonumber\\
    &= 2\, \mathrm{Pe}^{2} \int_0^t (t-\tau) C(\tau) \,d\tau,
    \label{eq:s-msd}
\end{align}
where
\begin{equation}
    C(\tau) = \left\langle \cos\!\left[ \theta(t+\tau)-\theta(t) \right] \right\rangle
\end{equation}
is the stationary orientational correlation function.
At long times, the MSD grows diffusively,
\begin{equation}
    \mathrm{MSD}(t) \simeq 4D_{\mathrm{eff}}t,
    \label{eq:msd-gg}
\end{equation}
from which the effective diffusion coefficient follows as
\begin{equation}
    D_{\mathrm{eff}} = \frac12 \mathrm{Pe}^2 \tau_\theta,
    \label{eq:deff}
\end{equation}
where
\begin{equation}
    \tau_\theta = \int_0^\infty C(\tau)\,d\tau
\end{equation}
is the integral correlation time.
Equation~\eqref{eq:deff} establishes a direct connection between orientational relaxation and long-time transport.
The remainder of this work is therefore devoted to understanding how different smectic textures modify the orientational dynamics, and consequently the effective diffusion coefficient.

\section{Planar Texture}
\label{sec:Planar}

We begin with the simplest smectic configuration, in which the layer-normal field is spatially uniform.
Choosing the layer normal to point along the $y$ direction [Fig.~\ref{fig:Illustration}(b)] gives $\theta_n=\pi/2$, so that the relative angle $\varphi(t)=\theta(t)-\pi/2$ depends only on the particle orientation and not on its position.
As a result, the orientational dynamics decouples from the translational motion and can be solved independently.
This geometry therefore isolates the effects of continuous alignment with the smectic layers, providing a natural reference against which the influence of curved and defect-laden textures can later be assessed. In particular, here $\tau_\theta$ does not depend on Pe, and the Péclet number enters the translational dynamics only as an overall prefactor, implying that the MSD scales as $\mathrm{Pe}^2$.
Throughout this section we measure length in units of the intrinsic persistence length, $L=l_p$, corresponding to $\mathrm{Pe}=1$ in Eq.~\eqref{eq:em-tpara}.
We nevertheless retain the explicit dependence on $\mathrm{Pe}$ throughout the analytical calculations in order to make this scaling transparent.

As a reference, we first consider the uncoupled limit $\varepsilon=0$, for which Eq.~\eqref{eq:p-aem} reduces to the orientational dynamics of the conventional active Brownian particle (ABP)~\cite{bechinger2016active}.
In this case, the orientational correlation function is obtained exactly as $C(\tau)=e^{-\tau}$, yielding an integral correlation time $\tau_\theta=1$ and a rotational mean-squared displacement
\begin{equation}
    \mathrm{RMSD}(t)=2t.
\end{equation}
Substituting $C(\tau)$ into Eq.~\eqref{eq:s-msd} gives
\begin{equation}
    \mathrm{MSD}(t)
    =
    2\left(t-1+e^{-t}\right)\mathrm{Pe}^2,
    \label{eq:s-abp}
\end{equation}
whose asymptotic behavior is
\begin{equation}
    \mathrm{MSD}(t)
    \approx
    \begin{cases}
        \mathrm{Pe}^2 t^2,
            & t\ll1,\\[1mm]
        2\,\mathrm{Pe}^2\,t,
            & t\gg1.
    \end{cases}
    \label{eq:ABPAsymptotic}
\end{equation}
The long-time limit recovers the well-known effective diffusion coefficient $D_\mathrm{eff} (\varepsilon=0) = (1/2)\,\mathrm{Pe}^2$.
Although $D_{\mathrm{eff}}$ is determined by the integral correlation time $\tau_\theta$, it is also useful to introduce an \emph{effective persistence time} $\tau_{\mathrm{eff}}$, defined as the characteristic duration of the ballistic regime before the crossover to diffusive motion.
For the conventional ABP, $\tau_{\mathrm{eff}}=2$ in our dimensionless units (or $2D_R^{-1}$ in physical units).
As we shall see, coupling to the smectic layers increases both $\tau_{\mathrm{eff}}$ and the long-time diffusivity.
The effective persistence time will play a central role in our analysis of transport through focal-conic textures, where it determines the time window over which particle trajectories can be approximated as ballistic between successive scattering events.

We now turn to the opposite limit of strong anchoring, $\varepsilon\gg1$, in which the particle orientation remains tightly confined around one of the minima of the alignment potential.
Choosing the minimum at $\theta=\pi/2$, Eq.~\eqref{eq:p-aem} can be linearized as
\begin{equation}
    \frac{d\varphi}{dt} 
        \approx -\varepsilon\,\varphi(t) + \sqrt{2}\,\xi(t),
    \label{eq:BHOEq}
\end{equation}
where $\varphi=\theta-\pi/2$. This equation describes an Ornstein--Uhlenbeck process, or equivalently a one-dimensional Brownian harmonic oscillator.
After the initial transient has decayed, the process reaches a stationary state. Assuming stationarity, Eq.~\eqref{eq:BHOEq} has the formal solution
\begin{equation}
    \varphi(t)
        = \sqrt{2} \int_{-\infty}^{t} e^{-\varepsilon(t-s)} \xi(s)\,ds,
\end{equation}
from which the rotational mean-squared displacement follows as
\begin{align}
    \mathrm{RMSD}(t)
        &= \left\langle [\Delta\theta(t)]^2 \right\rangle
        = \left\langle [\Delta\varphi(t)]^2 \right\rangle
        \nonumber\\
        &= \frac{2}{\varepsilon} \left(1-e^{-\varepsilon t} \right)
        \approx
        \begin{cases}
            2t, & \varepsilon \, t\ll1,\\[1mm]
            2/\varepsilon, & \varepsilon \, t\gg1.
    \end{cases}
    \label{eq:RMSD_Planar}
\end{align}
Since the orientational fluctuations remain small, the angular correlation function can be approximated by $C(\tau) \approx 1-\mathrm{RMSD}(\tau)/2$, which, together with Eq.~\eqref{eq:s-msd}, yields $\mathrm{MSD}(t) \approx \mathrm{Pe}^2t^2$, up to corrections of order $\varepsilon^{-1}$.
Thus, within the harmonic approximation the particle remains effectively ballistic over the entire time window accessible to the approximation.
The absence of a crossover to diffusive motion is not a physical prediction, but rather a limitation of the harmonic approximation itself.
By confining the dynamics to a single minimum of the alignment potential, Eq.~\eqref{eq:BHOEq} excludes the rare barrier-crossing events \cite{escape1,escape2,escape3} that eventually randomize the particle orientation.
As we show next, these activated transitions restore long-time diffusion and determine the effective persistence time in the strong-anchoring regime.

The harmonic approximation captures the local dynamics within a single potential well but cannot describe the rare reorientation events responsible for long-time transport.
These events correspond to thermally activated crossings between the two degenerate minima of the alignment potential, $ E(\varphi) = -(\varepsilon / 4) \cos(2\varphi)$, located at $\varphi=0$ and $\pi$.
In the strong-anchoring regime ($\varepsilon\gg1$), the particle therefore performs long nearly ballistic runs interrupted by infrequent orientation reversals, suggesting an effective one-dimensional run-and-tumble description.
The reversal rate is determined by thermal activation over the barriers located at $\varphi=\pm\pi/2$.
Since the potential has barrier height $\Delta E=\varepsilon/2$ and curvatures $E''(0)=|E''(\pi/2)|=\varepsilon$, Kramers' escape-rate theory~\cite{Kramers1940,HanggiBorkovec1990} predicts the rate of reorientation events.

Applying the overdamped Kramers formula to Eq.~\eqref{eq:p-aem} yields the reversal rate
\begin{equation}
    \Gamma(\varepsilon)
        = 2\, \frac{\sqrt{E''(0)\,|E''(\pi/2)|}}{2\pi} e^{-\Delta E}
        = \frac{\varepsilon}{\pi} e^{-\varepsilon/2},
    \label{eq:Gamma-total}
\end{equation}
where the prefactor of two accounts for escape through either barrier at $\varphi=\pm\pi/2$.
The corresponding mean waiting time between successive reversals is $\Gamma^{-1}$.
The exact mean-squared displacement of a one-dimensional run-and-tumble particle~\cite{Weiss2002,SantraGupta2024} with reversal rate $\Gamma$ yields a diffusion coefficient $\mathrm{Pe}^2/(2\Gamma)$ in our units.
Throughout this work, however, we define the effective diffusion coefficient through the two-dimensional convention $\mathrm{MSD}(t)=4D_{\mathrm{eff}}t$.
Consequently, the corresponding diffusion coefficient is
\begin{equation}
    D_{\mathrm{eff}}
        = \frac{\mathrm{Pe}^2}{4\,\Gamma}
        = \frac{\pi\,\mathrm{Pe}^2}{4\,\varepsilon} e^{\varepsilon/2}.
    \label{eq:deff-kramers}
\end{equation}
Matching the short- and long-time asymptotic behaviors of the exact run-and-tumble mean-squared displacement further shows that the ballistic-to-diffusive crossover occurs after approximately one quarter of the mean waiting time between successive reversals.
This motivates the definition of the effective persistence time,
\begin{equation}
    \tau_{\mathrm{eff}}
        = \frac{1}{4\, \Gamma}
        = \frac{\pi}{4\, \varepsilon} e^{\varepsilon/2},
    \label{eq:tauEffKramers}
\end{equation}
which grows exponentially with the anchoring strength, reflecting the increasingly rare orientational reversals.
As we show below, Eqs.~\eqref{eq:deff-kramers} and~\eqref{eq:tauEffKramers} accurately describe the numerical simulations in the strong-anchoring regime.
More importantly, the effective persistence time $\tau_{\mathrm{eff}}$ provides the central physical quantity governing transport through periodic arrays of focal-conic domains, where it competes directly with the mean time between successive encounters with smectic defects.

\begin{figure}[!h]
\centering
\includegraphics[width=1.0\linewidth]{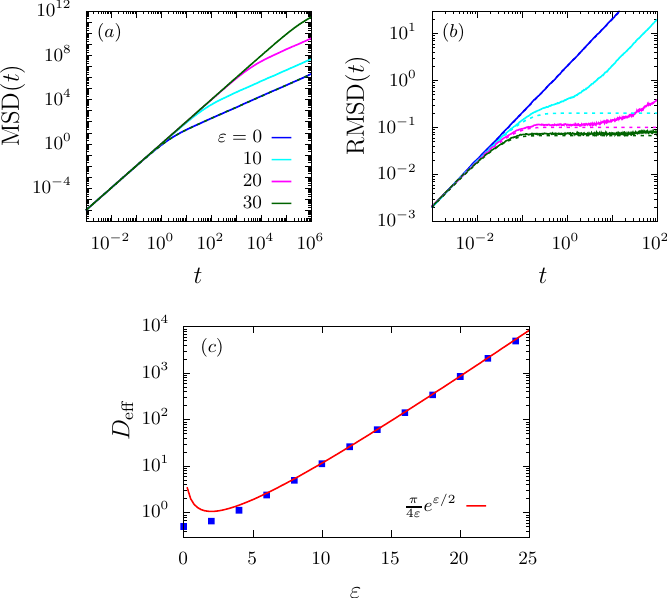}
\caption{Transport properties of an active particle moving in a planar Sm-A texture for different anchoring strengths $\varepsilon$. (a) Mean-squared displacement (MSD) and (b) rotational mean-squared displacement (RMSD) as functions of time. Solid lines show numerical simulations of the full equations of motion, while dashed lines correspond to the strong-anchoring harmonic approximation. (c) Effective diffusion coefficient $D_{\mathrm{eff}}$ obtained from simulations (blue squares) compared with the strong-anchoring Kramers prediction, Eq.~\eqref{eq:deff-kramers} (red line).}
\label{fig:MSD_Planar}
\end{figure}

Figure~\ref{fig:MSD_Planar} summarizes the transport properties of an active particle in a planar smectic texture and compares the numerical results with the analytical approximations developed above.
Figures~\ref{fig:MSD_Planar}(a) and~\ref{fig:MSD_Planar}(b) show, respectively, the mean-squared displacement (MSD) and the rotational mean-squared displacement (RMSD) as functions of time for different values of the anchoring strength $\varepsilon$.
Solid lines correspond to numerical integration of the full equations of motion, while dashed lines show the predictions of the harmonic approximation, namely $\mathrm{MSD}(t)\simeq \mathrm{Pe}^2t^2$ in Fig.~\ref{fig:MSD_Planar}(a) and Eq.~\eqref{eq:RMSD_Planar} in Fig.~\ref{fig:MSD_Planar}(b).
As expected, the harmonic approximation accurately describes the dynamics while the particle remains confined near a single minimum of the alignment potential.
At sufficiently long times, rare transitions between the two orientational minima become relevant, leading to the crossover from ballistic to diffusive motion that is absent from the harmonic description.
This crossover is increasingly delayed as $\varepsilon$ increases, reflecting the strong enhancement of the effective persistence time.
Figure~\ref{fig:MSD_Planar}(c) provides a quantitative test of the long-time theory by comparing the effective diffusion coefficient extracted from simulations (blue squares) with the strong-anchoring prediction of Eq.~\eqref{eq:deff-kramers} (red line).
The Kramers-based approximation captures the rapid increase of $D_{\mathrm{eff}}$ with $\varepsilon$ and shows good quantitative agreement with the simulations already for moderate anchoring strengths, $\varepsilon\gtrsim5$.
Together, these results establish the planar texture as a controlled reference problem: local orientational fluctuations are described by the harmonic approximation, whereas rare activated reversals determine the effective persistence and long-time diffusion.
We next examine how this picture is modified by spatially varying layer orientation and singular defects, beginning with an isolated toroidal focal conic domain.

\section{One Toroidal Focal Conic Domain}
\label{sec:OneFCD}

\begin{figure}[!h]
\centering
\includegraphics[width=1.0\linewidth]{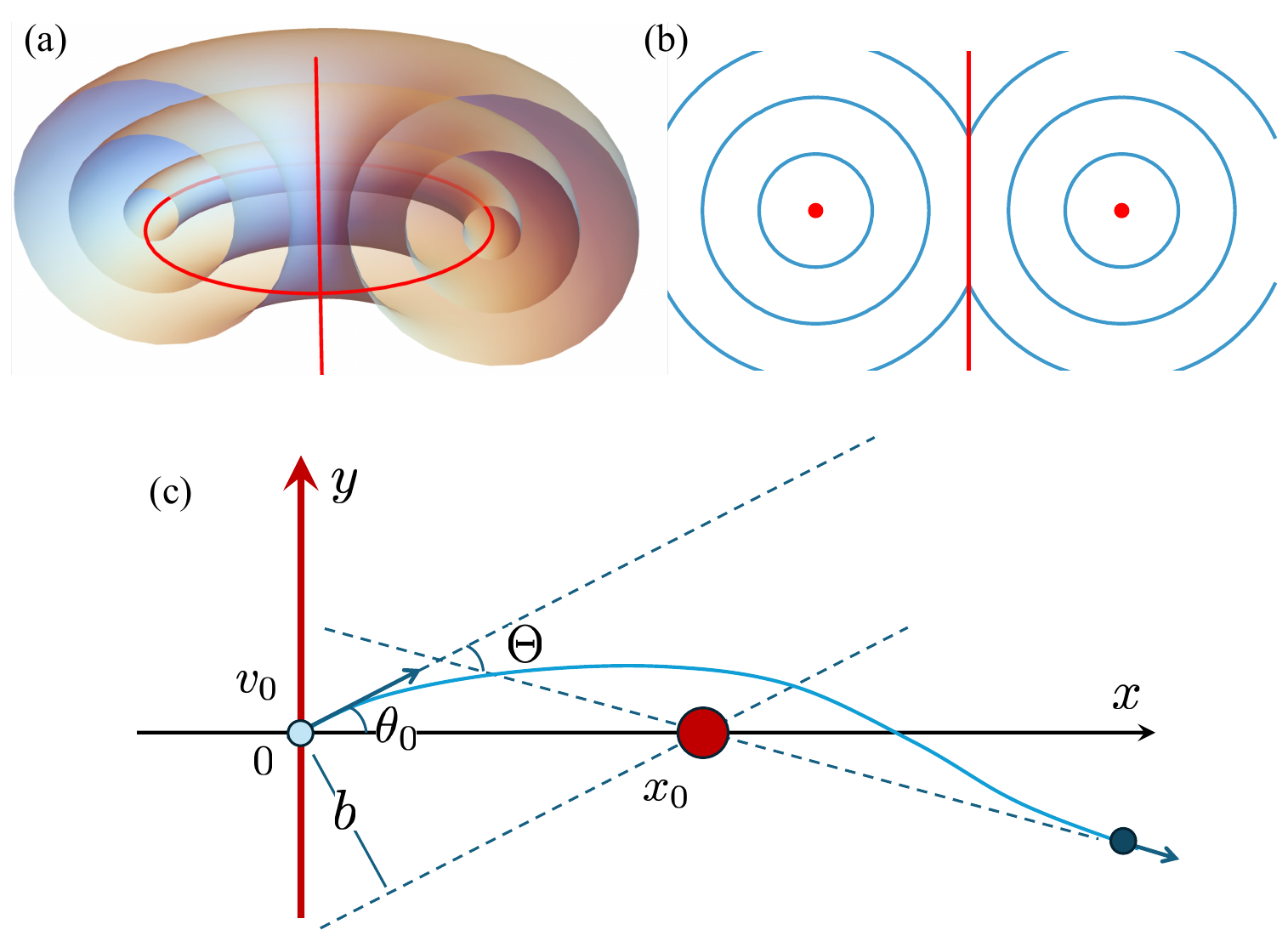}
\caption{Geometry and scattering in a single toroidal focal conic domain (TFCD). (a) Smectic layers (light blue surfaces) and focal singularities (red curves), consisting of a circular focal line and a straight line passing through its center and perpendicular to its plane. (b) Two-dimensional section containing the straight focal line; the circular focal line intersects the section at two point defects. (c) Schematic representation of a scattering trajectory launched from the origin at an initial angle $\theta_0$ with respect to the $x$ axis. The defect is located at $(x_0,0)$.}
\label{fig:TFCD}
\end{figure}

We next introduce spatial variations of the smectic texture by considering a single \emph{toroidal focal conic domain} (TFCD)~\cite{kleman2003ch10}.
In this geometry, space is filled by a family of smooth smectic layers whose normals converge toward a pair of focal sets: a circular singular line and a straight singular line passing through its center and perpendicular to its plane [Fig.~\ref{fig:TFCD}(a)].
The layer-normal field is therefore well defined away from these focal sets but becomes singular on them. 
For sufficiently strong alignment, the curved layers continuously steer active particles along the local layer normal and toward the focal sets.
Encounters with these singularities interrupt this smooth guidance and can produce abrupt changes in the direction of motion, providing an elementary scattering mechanism that is absent in the planar texture.
We restrict our analysis to the two-dimensional section containing the straight focal line shown in Fig.~\ref{fig:TFCD}(b).
In this section, the circular focal line intersects the plane at two point defects located symmetrically about the straight singular line.
This choice provides a particularly simple geometry in which individual defect encounters can be formulated as a conventional scattering problem.
We emphasize that the periodic array considered in Sec.~\ref{sec:ArrayFCDs} will be analyzed using a different two-dimensional section, containing the circular focal lines of the TFCDs.
Consequently, the singular sets encountered by the particles will have a different geometry.
The present calculation should therefore be viewed as a characterization of the elementary mechanism by which focal singularities reorient active trajectories, rather than as a direct microscopic description of the scattering events in the array.
We first determine how the deflection produced by an isolated focal defect depends on the alignment strength $\varepsilon$.
We then examine the resulting translational and orientational dynamics, identifying the characteristic signatures of defect encounters and their scaling with the particle speed.

The scattering geometry is illustrated in Fig.~\ref{fig:TFCD}(c).
An active particle is launched from the origin with speed $v$ and initial orientation $\theta_0$ relative to the $x$ axis.
We restrict the initial angle to $0<\theta_0<\pi/4$.
Trajectories with $\theta_0<0$ are related by reflection symmetry, whereas for $\theta_0>\pi/4$ the particle is instead attracted toward the straight focal line at $x=0$ and remains confined in its vicinity.
To characterize the encounter with the point defect, we introduce an effective impact parameter $b=x_0\sin\theta_0$, where $x_0$ is the distance between the point defect and the straight focal line.
This quantity plays a role analogous to the impact parameter of conventional scattering theory: it measures the transverse distance between the unperturbed trajectory and the scattering center.
The analogy is not exact, however, because the smectic director field extends throughout space and continuously acts on the particle rather than defining a localized scattering potential.

For this calculation we consider the deterministic equations of motion, setting the noise term $\xi$ in Eqs.~\eqref{eq:em-tpara} and~\eqref{eq:p-aem} to zero.
We choose $L=x_0$ as the unit of length, such that $\mathrm{Pe}=v\tau_R / x_0$.
In the absence of rotational noise, $\tau_R$ simply fixes the unit of time, and the deterministic scattering dynamics is controlled by the ratio
\begin{equation}
    \lambda\equiv\frac{\varepsilon}{\mathrm{Pe}},
\end{equation}
which compares the rate of alignment with the rate at which the particle traverses the texture.
We now estimate the deflection angle $\Theta$ and the corresponding two-dimensional differential cross section in the weak-coupling regime, $\lambda\ll1$.

In this regime, the orientational change during the encounter is small and the trajectory can, to leading order, be approximated by the unperturbed straight path with $\theta(t)\simeq\theta_0$.
This procedure is analogous to a first Born approximation: the orientational response is evaluated along the unperturbed trajectory and used to obtain the leading correction to the motion.
Evaluating the orientational torque along this trajectory and integrating Eqs.~\eqref{eq:em-tpara} and~\eqref{eq:p-aem} gives
\begin{equation}
    \Theta \equiv \theta(\Lambda)-\theta_0
    = \int_0^\Lambda \lambda\, b\,\frac{s-a}{(s-a)^2+b^2}\,ds,
    \label{eq:deflection-integral}
\end{equation}
where $s=\mathrm{Pe}\,t$, $a=\cos\theta_0$, and $b=\sin\theta_0$.
The integral grows logarithmically at large $s$.
This apparent divergence is an artifact of extending the straight-trajectory approximation to arbitrarily long times: Even for weak coupling, the accumulated torque eventually produces an appreciable reorientation and invalidates the assumption $\theta(t)\simeq\theta_0$.
We therefore introduce a self-consistent cutoff at the distance over which alignment produces an order-one orientational change.
We set $\Lambda\sim \mathrm{Pe}/\varepsilon =\lambda^{-1}$, so that Eq.~\eqref{eq:deflection-integral} yields
\begin{align}
    \Theta &=
    \frac{\lambda \, b}{2} \ln \left[(\Lambda-\cos\theta_0)^2+\sin^2\theta_0\right]
    \nonumber\\
    &\simeq
    \lambda
    \sin\theta_0
    \ln \left(\frac{1}{\lambda}\right),
    \label{eq:deflection-weak}
\end{align}
where the second expression retains the leading behavior for $\lambda\ll1$.
The corresponding two-dimensional differential cross section is therefore
\begin{equation}
    \frac{d\sigma}{d\Theta} = \left|\frac{db}{d\Theta}\right|
    \simeq \frac{1}
         {\lambda \ln(1/\lambda)},
    \label{eq:cross-section-weak}
\end{equation}
within the small-angle approximation, for which $b=\sin\theta_0\simeq\theta_0$.

Several features of this estimate have a simple physical interpretation.
A trajectory directed exactly toward the defect, $\theta_0=0$, has zero deflection by symmetry. 
More generally, Eq.~\eqref{eq:deflection-weak} shows that the scattering strength is governed by the ratio $\lambda=\varepsilon/\mathrm{Pe}$: Increasing the alignment strength or decreasing the propulsion speed enhances the deflection.
Conversely, Eq.~\eqref{eq:cross-section-weak} shows that weak deflections occupy an increasingly large range of impact parameters as $\lambda$ decreases.
The logarithmic factor should not, however, be regarded as a quantitatively controlled prediction, since its precise form depends on the self-consistent cutoff $\Lambda \sim \lambda^{-1}$.
Also, notice that the straight-trajectory approximation eventually breaks down when the accumulated deflection is no longer small, parametrically when $\lambda b\gtrsim1$.
A more complete treatment could be obtained by analyzing the coupled deterministic equations as a nonlinear dynamical system, but this lies beyond the scope of the present work.

\begin{figure}[!h]
    \centering
    \includegraphics[width=1.0\linewidth]{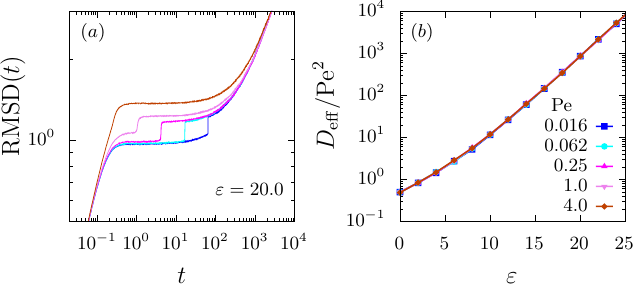}
    \caption{Transport in a single toroidal focal conic domain. (a) Rotational mean-squared displacement (RMSD) as a function of time for fixed anchoring strength $\varepsilon=20$ and several values of $\mathrm{Pe}$. Sharp intermediate-time steps arise from rapid reorientations during encounters with focal defects. (b) Rescaled effective diffusion coefficient $D_{\mathrm{eff}}/\mathrm{Pe}^2$ as a function of $\varepsilon$ for the same values of $\mathrm{Pe}$. The collapse of the curves shows that an isolated TFCD preserves the asymptotic scaling $D_{\mathrm{eff}}\sim\mathrm{Pe}^2$.}
    \label{fig:Diff1FCD}
\end{figure}

We conclude this section by examining the transport properties obtained from the full stochastic equations of motion, Eqs.~\eqref{eq:em-tpara} and~\eqref{eq:p-aem}.
The numerical protocol is similar to that used in the scattering analysis: Particles are launched from the origin with an initial orientation $\theta_0$ relative to the $x$ axis, but we now sample $\theta_0$ without restricting it to the scattering interval considered above. 
Figure~\ref{fig:Diff1FCD}(a) shows the rotational mean-squared displacement as a function of time for fixed anchoring strength $\varepsilon=20$ and $\mathrm{Pe}=1/64,\,1/16,\,1/4,\,1$, and $4$.
The overall behavior resembles that found for the planar texture: an initial growth is followed by an intermediate plateau and, ultimately, by the long-time regime associated with rotational diffusion.
The presence of the focal defects, however, leaves a distinct signature in the intermediate-time dynamics.
For appropriate values of $\mathrm{Pe}$, the $\mathrm{RMSD}$ exhibits sharp steps corresponding to rapid re-orientations produced by encounters with the singularities.
These events occur at a characteristic time
\begin{equation}
    \tau_{\mathrm{d}}\sim\mathrm{Pe}^{-1},
\end{equation}
which is the time required for a particle to traverse a distance of order the TFCD size and reach a focal defect.
The visibility of these steps is controlled by the competition between this geometric traversal time and the intrinsic orientational dynamics.
At large $\mathrm{Pe}$, defect encounters occur so rapidly that their signatures merge with the initial angular relaxation.
At small $\mathrm{Pe}$, by contrast, the traversal time becomes sufficiently long that rotational diffusion substantially de-correlates the particle orientation before the characteristic encounter, washing out the sharp scattering signature.
The steps are therefore most clearly resolved at intermediate values of $\mathrm{Pe}$.

Despite these pronounced transient effects, an isolated TFCD does not modify the asymptotic scaling of the translational transport.
Figure~\ref{fig:Diff1FCD}(b) shows the rescaled effective diffusion coefficient $D_{\mathrm{eff}}/\mathrm{Pe}^2$ as a function of $\varepsilon$ for the same values of $\mathrm{Pe}$.
The collapse of the different curves demonstrates that $D_{\mathrm{eff}}\sim\mathrm{Pe}^2$, as in the planar texture.
This behavior has a simple physical origin.
After interacting with the focal region, most particles eventually escape and subsequently revisit it only sporadically.
Consequently, a single TFCD can strongly modify individual trajectories and produce clear intermediate-time signatures, but it does not introduce a persistent sequence of scattering events capable of controlling the long-time dynamics.
This observation motivates the geometry considered in the next section.
By arranging TFCDs periodically, particles encounter focal defects repeatedly rather than only transiently.
As we shall see, these repeated scattering events introduce a geometry-controlled persistence mechanism that competes with the intrinsic persistence of the active particle and qualitatively changes the scaling of long-time transport.

\section{Periodic Arrays of Toroidal Focal Conic Domains}
\label{sec:ArrayFCDs}

\begin{figure}[t]
    \centering
    \begin{minipage}[t]{0.49\columnwidth}
        \raggedright\textbf{$(a)$}\\[2pt]
        \centering
        \includegraphics[width=\linewidth]{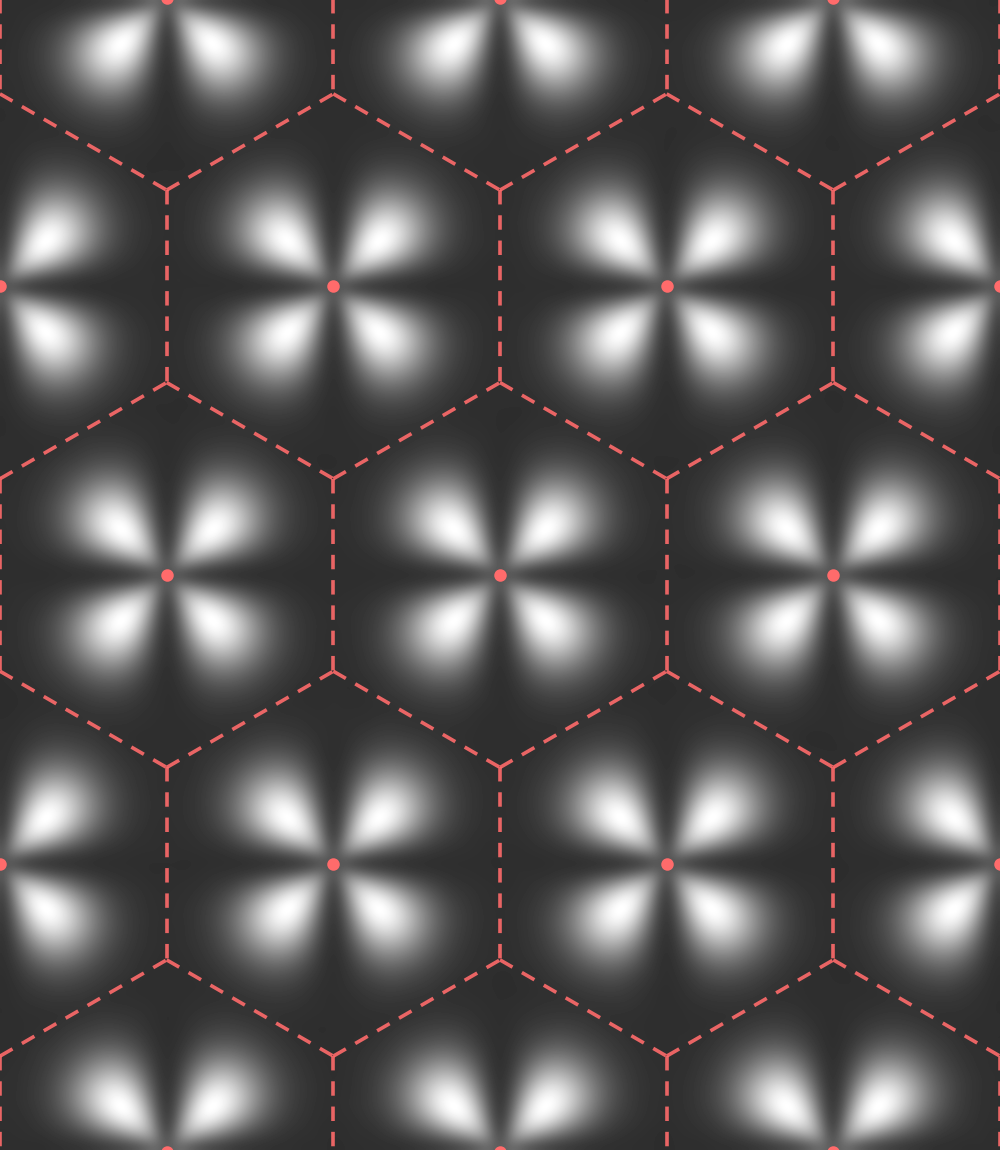}
    \end{minipage}\hfill
    \begin{minipage}[t]{0.49\columnwidth}
        \raggedright\textbf{$(b)$}\\[2pt]
        \centering
        \includegraphics[width=\linewidth]{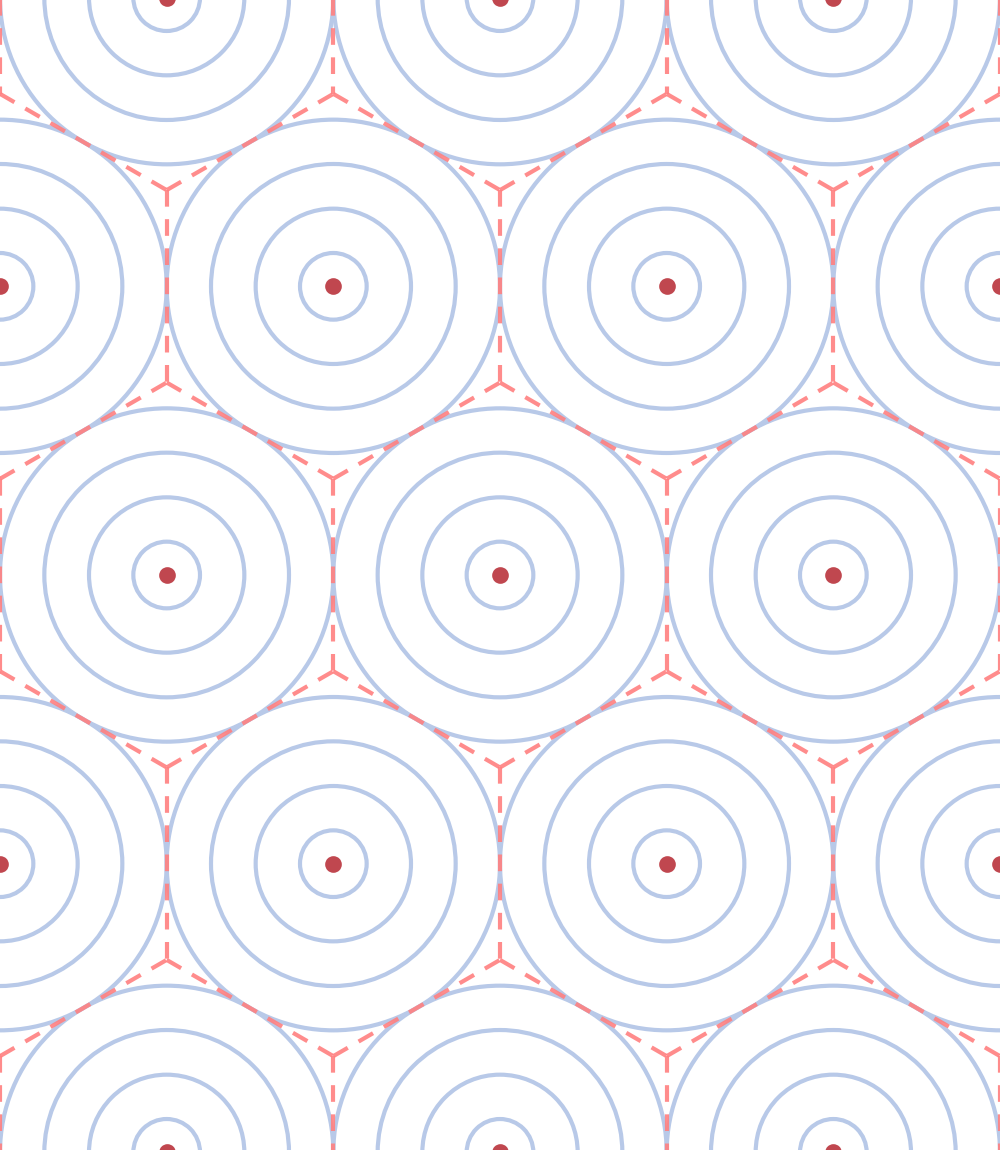}
    \end{minipage}
    \caption{Periodic array of toroidal focal conic domains.
    $(a)$ Computer-generated illustration of a crossed-polarizer microscopy image showing a texture with a triangular array of toroidal focal conic domains. 
    $(b)$ Schematic representation of the corresponding two-dimensional texture.
    Within each domain, the layer normals (radial lines) form a concentric pattern centered at the intersection with the straight focal line (red dots).
    In both panels, red dashed lines indicate the singular boundaries separating neighboring domains, which act as recurrent scatterers for active particles moving through the array.}
    \label{fig:Array_FCDs}
\end{figure}

We now turn to a periodic array of toroidal focal conic domains, in which encounters with smectic defects become a recurrent feature of the particle dynamics.
Figure~\ref{fig:Array_FCDs}$(a)$ shows a computer-generated illustration mimicking a crossed-polarizer microscopy image of a smectic section forming a triangular array of TFCDs, reproducing the characteristic four-lobed pattern observed experimentally (see Fig.~7 in Ref.~\cite{BlancKleman1999}). Unlike the isolated-TFCD geometry considered in Sec.~\ref{sec:OneFCD}, we now analyze the two-dimensional section containing the circular focal lines of the domains.
This is the natural section for describing the periodic texture and the transport of active particles across neighboring TFCDs.
Within each domain, the layer normals form the concentric pattern illustrated schematically in Fig.~\ref{fig:Array_FCDs}(b), centered at the point where the straight focal line intersects the section. 
Neighboring domains meet along singular boundaries [red dashed lines in Fig.~\ref{fig:Array_FCDs}(b)], across which the local layer-normal field changes abruptly.
These boundaries therefore act as recurrent scatterers for active particles moving through the array.
This situation differs qualitatively from the isolated TFCD: rather than eventually escaping the defect region and encountering it only sporadically, a particle traversing the array repeatedly reaches singular boundaries after traveling a distance set by the domain size.
Repeated scattering consequently introduces a geometric time scale that can compete with the intrinsic persistence of the active motion and, as we show below, modify the asymptotic scaling of transport.
We focus on the triangular arrangement of Fig.~\ref{fig:Array_FCDs} as a minimal model in which to investigate this competition.
The mechanism is not expected to depend on the particular lattice symmetry: analogous behavior may arise in other periodic smectic textures, including square arrays of parabolic focal conic domains~\cite{ChatterjeeAnna2012} and more complex periodic patterns observed in smectic thin films~\cite{ZapponeLacaze2012}.
Our objective is therefore not to characterize the detailed transport properties of a particular focal-conic lattice, but to identify the dynamical consequences of repeated defect scattering and the resulting competition between intrinsic and geometry-induced persistence.
The long-time transport through the TFCD array can be understood in terms of the competition between two characteristic time scales.
The first is the intrinsic effective persistence time $\tau_{\mathrm{eff}}$, which characterizes spontaneous reorientation of the active particle.
In Sec.~\ref{sec:Planar}, we used Kramers' escape-rate theory to obtain an analytical estimate of $\tau_{\mathrm{eff}}$ in the strong-anchoring regime [Eq.~\eqref{eq:tauEffKramers}].
The second is the geometric scattering time $\tau_d$, defined as the characteristic time between successive encounters with singular boundaries.
As discussed for the isolated TFCD in Sec.~\ref{sec:OneFCD}, this time is set by the ratio between the characteristic distance between defects and the particle speed.
In the dimensionless units used here, $\tau_d \sim \mathrm{Pe}^{-1}$.
Both mechanisms can reorient the direction of motion.
We therefore expect the shorter of these two time scales to control the persistence of the trajectory,
\begin{equation}
    \tau \sim \min\left(\tau_{\mathrm{eff}},\tau_d\right),
    \label{eq:dominantPersistence}
\end{equation}
provided that an encounter with a singular boundary produces an appreciable reorientation.

A simple kinetic estimate then predicts the scaling of the long-time diffusion coefficient.
For persistent motion with characteristic reorientation time $\tau$, dimensional considerations give
\begin{equation}
    D_{\mathrm{eff}}
    \sim
    \mathrm{Pe}^2\,\tau.
    \label{eq:ArrayKineticEstimate}
\end{equation}
Consider first the regime $\tau_{\mathrm{eff}}\ll\tau_d$.
The particle typically loses its intrinsic orientational persistence before reaching the next singular boundary, so that defect scattering does not control the long-time dynamics.
In this case,
\begin{equation}
    D_{\mathrm{eff}}
    \sim
    \mathrm{Pe}^2\tau_{\mathrm{eff}}(\varepsilon)
    \propto
    \mathrm{Pe}^2
    \qquad
    (\tau_{\mathrm{eff}}\ll\tau_d),
    \label{eq:ArrayPe2}
\end{equation}
at fixed $\varepsilon$. For $\varepsilon\ll1$, this reduces to the conventional ABP scaling, while at strong anchoring the prefactor is enhanced by the increase of $\tau_{\mathrm{eff}}$ derived in Sec.~\ref{sec:Planar}.
A qualitatively different regime emerges when $\tau_d\ll\tau_{\mathrm{eff}}$.
The particle then remains intrinsically persistent over the distance separating neighboring defects, and its direction of motion is instead randomized primarily by repeated encounters with the singular boundaries.
The relevant persistence time is consequently set by the geometry, $\tau\sim\tau_d\sim\mathrm{Pe}^{-1}$.
Substitution into Eq.~\eqref{eq:ArrayKineticEstimate} immediately gives
\begin{equation}
    D_{\mathrm{eff}}
    \sim
    \mathrm{Pe},
    \qquad
    (\tau_d\ll\tau_{\mathrm{eff}}).
    \label{eq:ArrayPe1}
\end{equation}
We therefore predict a crossover from the conventional $D_{\mathrm{eff}}\sim\mathrm{Pe}^2$ scaling to a defect-mediated regime with $D_{\mathrm{eff}}\sim\mathrm{Pe}$.
In the latter regime, the effective persistence is no longer determined primarily by the intrinsic orientational dynamics of the active particle, but emerges from the characteristic distance and time between successive scattering events in the smectic texture.

This simple argument assumes that each encounter with a singular boundary produces a significant reorientation and is therefore expected to apply only over a finite range of particle speeds.
At sufficiently large $\mathrm{Pe}$, the traversal time across the defect region becomes too short for the alignment torque to generate an appreciable angular deflection.
Defect encounters then cease to act as effective reorientation events, and intrinsic orientational relaxation again controls the long-time transport.
We consequently expect a return toward the conventional $D_{\mathrm{eff}}\sim\mathrm{Pe}^2$ scaling at sufficiently large $\mathrm{Pe}$.

\begin{figure}[!h]
    \centering
    \includegraphics[width=.9\linewidth]{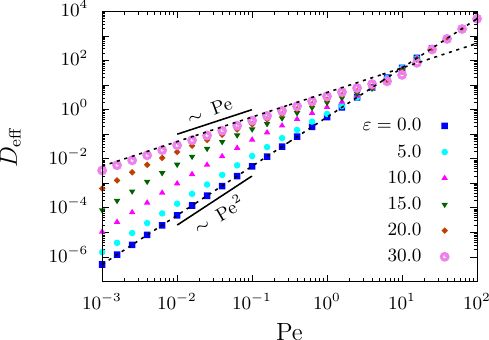}
    \caption{Long-time effective diffusion coefficient $D_{\mathrm{eff}}$ as a function of the Péclet number $\mathrm{Pe}$ for several values of the anchoring strength $\varepsilon$.
    Increasing $\varepsilon$ produces a crossover from the conventional $D_{\mathrm{eff}}\sim\mathrm{Pe}^2$ scaling to the defect-mediated $D_{\mathrm{eff}}\sim\mathrm{Pe}$ regime at small to intermediate $\mathrm{Pe}$.
    At sufficiently large $\mathrm{Pe}$, rapid traversal of the defect regions suppresses effective scattering and the conventional scaling is progressively recovered.
    Reference slopes proportional to $\mathrm{Pe}$ and $\mathrm{Pe}^2$ are shown as guides to the eye.}
    \label{fig:DiffArray}
\end{figure}

Figure~\ref{fig:DiffArray} shows the effective diffusion coefficient $D_{\mathrm{eff}}$ as a function of the Péclet number $\mathrm{Pe}$ for several values of the anchoring strength $\varepsilon$.
In agreement with the scaling argument above, at small to intermediate $\mathrm{Pe}$ we observe a clear crossover from the conventional $D_{\mathrm{eff}}\sim\mathrm{Pe}^2$ behavior at weak anchoring, where $\tau_{\mathrm{eff}}\ll\tau_d$, to the defect-mediated regime $D_{\mathrm{eff}}\sim\mathrm{Pe}$ at strong anchoring, where $\tau_d\ll\tau_{\mathrm{eff}}$.
At sufficiently large $\mathrm{Pe}$, however, the defect-mediated scaling breaks down: Particles traverse the singular regions too rapidly for the alignment torque to produce substantial reorientation, and the transport progressively returns toward the intrinsic $D_{\mathrm{eff}}\sim\mathrm{Pe}^2$ scaling.

\begin{figure}[!h]
    \centering
    \includegraphics[width=.9\linewidth]{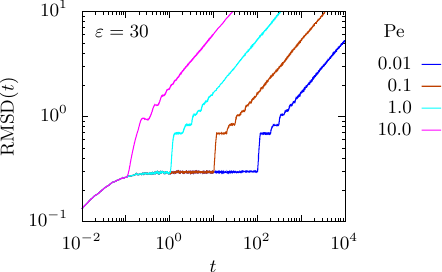}
    \caption{Rotational mean-squared displacement (RMSD) as a function of time for a periodic array of toroidal focal conic domains, at fixed anchoring strength $\varepsilon=30$ and several values of $\mathrm{Pe}$.
    The successive steps correspond to reorientation events induced by scattering at the singular boundaries between neighboring domains.
    Their characteristic temporal separation scales as $\tau_d\sim\mathrm{Pe}^{-1}$, providing a dynamical signature of the repeated-scattering mechanism responsible for the $D_{\mathrm{eff}}\sim\mathrm{Pe}$ regime.}
    \label{fig:RMSDArray}
\end{figure}

We conclude this section by examining the rotational mean-squared displacement (RMSD), which provides direct evidence of the repeated-scattering mechanism underlying the $D_{\mathrm{eff}}\sim\mathrm{Pe}$ scaling.
Figure~\ref{fig:RMSDArray} shows $\mathrm{RMSD}(t)$ for fixed strong anchoring, $\varepsilon=30$, and several values of $\mathrm{Pe}$.
In contrast to the isolated TFCD, where defect encounters produce occasional abrupt reorientations, the periodic array gives rise to a sequence of distinct steps in the RMSD.
These steps reflect successive scattering events at the singular boundaries separating neighboring domains, with a characteristic time interval $\tau_d\sim\mathrm{Pe}^{-1}$.
At longer times, the individual steps become less distinguishable, giving way to an approximately linear growth of the RMSD associated with rotational diffusion.
This behavior provides a direct dynamical signature of the geometry-controlled reorientation process: repeated encounters with focal defects establish a persistence time proportional to $\mathrm{Pe}^{-1}$, leading to the long-time transport scaling $D_{\mathrm{eff}}\sim\mathrm{Pe}$.

\section{Conclusions}
\label{sec:Conclusions}

In this work, we have investigated the transport of active particles through spatially structured anisotropic environments, using smectic liquid-crystal textures as a model system.
We considered three representative geometries: a planar texture with uniformly spaced layers, an isolated toroidal focal conic domain, and a periodic array of such domains.
In the planar configuration, alignment with the local layer normal enhances orientational persistence by suppressing spontaneous reversals.
Introducing a single focal conic domain reveals an additional mechanism of reorientation through scattering at geometrical singularities, which produces distinct signatures in the particle dynamics without altering the conventional long-time scaling $D_{\mathrm{eff}}\sim\mathrm{Pe}^2$.
In periodic arrays, however, repeated encounters with focal defects introduce a geometry-controlled scattering time that competes with the intrinsic persistence of active motion, giving rise to a distinct transport regime characterized by $D_{\mathrm{eff}}\sim\mathrm{Pe}$.
Our results thus demonstrate how the interplay between intrinsic orientational dynamics and the geometry of an anisotropic environment can generate emergent persistence mechanisms and qualitatively modify long-time transport.

An experimental realization of the mechanisms described here requires a suitable combination of several length scales, including the active-particle size, smectic layer spacing, focal-conic domain size, and intrinsic persistence length.
Lyotropic smectics~\cite{BlancKleman1999} and colloidal smectics~\cite{Grelet2025} offer promising platforms for controlling the relationship between smectic ordering and the dimensions of active inclusions, although focal conic domains are not typically observed in the latter.
Bacteria interacting with thin smectic films~\cite{BadhaGharbi2024} provide a complementary biological setting for investigating active transport in anisotropic ordered environments, with potential applications to defect-based biosensing.
Establishing experimental conditions under which active particles experience sufficiently strong alignment and recurrent scattering at focal defects remains an important challenge.

We conclude by discussing some limitations of our model that also suggest directions for future investigation.
Our description is restricted to two dimensions and employs a prescribed layer-normal field, without explicitly resolving individual smectic layers or their elastic deformations.
Although the selected two-dimensional sections represent preferred manifolds for particle motion, in the sense that particles confined to these sections tend to remain there in the absence of sufficiently strong perturbations, a fully three-dimensional treatment would allow trajectories to explore the complete focal-conic geometry.
We also neglect hydrodynamic interactions \cite{Menzel}, back-action on the smectic texture, and interactions between
active particles, all of which may become relevant in experimental realizations.
Finally, we assume preferential alignment of the propulsion direction with the local layer normal.
Alternative anchoring conditions, including alignment parallel to the layers, could lead to qualitatively different transport behavior.
Exploring these extensions may reveal additional mechanisms through which the geometry and topology of anisotropic environments influence active motion.

	\acknowledgments

    We thank Christophe Blanc and Eric Grelet for useful comments and suggestions.
    Y.M.A. acknowledges support from the Brazilian National Council for Scientific and Technological Development (CNPq) through a doctoral fellowship.
    W.G.C.O. acknowledges support from São Paulo Research Foundation (FAPESP) under Grant No. 2024/23876-3.
    A. R. S. acknowledges support from CNPq through a master fellowship.
    H.L. was supported by the German Research Foundation (DFG) within project number 572589640.
    A.P.V. acknowledges support from FAPESP under grants 2022/15453-0 and 2023/12490-4.
    D.B.L. acknowledges support from FAPESP under Grants No. 2021/14285-3 and No. 2022/09615-7.

	\bibliography{refs}

\end{document}